\documentclass[twocolumn,fleqn,usenatbib]{mnras}

\usepackage{amsfonts}
\usepackage{amsmath}
\usepackage{amssymb}
\usepackage{aas_macros}
\usepackage{graphicx}
\usepackage{color}
\usepackage{float}
\usepackage{hyperref}
\usepackage{cleveref}
\usepackage{tikz}
\usetikzlibrary{decorations.markings}
\usepackage{gnuplot-lua-tikz}

\def\lsim{~\rlap{$<$}{\lower 1.0ex\hbox{$\sim$}}}
\def\bsim{~\rlap{$>$}{\lower 1.0ex\hbox{$\sim$}}}

\def\apjl{Astrophys.\ J.\ Lett.}

\def\aap{Astron.\ Astrophys.}

\def\apjs{Astrophys.\ J. Supp.}

\def\la{\langle}

\def\mathbi#1{\textbf{\em #1}}

\def\evh{\mathrm{\hat{\bf{e}}}}
\def\kvh{\mathrm{\hat{\bf{k}}}}

\def\rvh{\mathrm{\hat{\bf{r}}}}
\def\xvh{\mathrm{\hat{\bf{x}}}}
\def\yvh{\mathrm{\hat{\bf{y}}}}
\def\zvh{\mathrm{\hat{\bf{z}}}}

\def\vk{\mathbi{k}}

\def\vr{\mathbi{r}}

\def\vu{\mathbi{u}}
\def\vv{\mathbi{v}}

\def\tw{\tilde \omega}

\def\mach{\mathcal{M}}

\def\fdf{\mathbi{F}_\text{DF}}

\newcommand{\intw}{\int_\omega}

\def\sllm{S_{\ell,\ell-1}^{m}}

\newcommand{\del}[1]{\delta^D\!\left(#1\right)}
\newcommand{\bessJb}[2]{J_{#1}\!\left(#2\right)}

\definecolor{red}{cmyk}{0,1,1,0.55}
\definecolor{blue}{rgb}{0.15, 0.2, .85}

\usepackage[normalem]{ulem}

\title[Orbital evolution of eccentric perturbers under dynamical friction]{Orbital evolution of eccentric perturbers under dynamical friction: crossing the sound barrier}

\author[Robin Buehler et al.]{Robin Buehler,\thanks{E-mail: \href{mailto:robinbuehler@campus.technion.ac.il}{robinbuehler@campus.technion.ac.il}}
Roman Kolyada,
and Vincent Desjacques
\\
Physics department, Technion, 3200003 Haifa, Israel}

\date{Accepted XXX. Received YYY; in original form ZZZ}

\pubyear{2023}

\begin{document}

\label{firstpage}
\pagerange{\pageref{firstpage}--\pageref{lastpage}}
\maketitle

\begin{abstract}
    In a gaseous medium, dynamical friction (DF) reaches a maximum when the orbital speed of a (point-like) perturber moving on a circular orbit is close to the sound speed. Therefore, in a quasi-steady state, eccentric orbits of perturbers approaching the sound barrier (from below) should rapidly circularize as they experience the strongest drag at pericenter passage. To investigate this effect, we extend the solution of \cite{desjacques/etal:2022} for circular DF in a uniform gaseous medium to eccentric Keplerian orbits. We derive an approximation to the steady-state DF force, which is valid for eccentricities as high as $e=0.9$ in a limited range of Mach number around the transition to supersonic regime. 
    We validate our analytical result with 3-dimensional simulations of the gas density response. Although gaseous DF generally dissipates orbital energy, we find that it can be directed along the motion of the perturber near pericenter passage when the eccentricity is $e\gtrsim 0.9$. 
    We apply our results to compute the long-time evolution of the orbital parameters. Most trajectories tend to circularize as the perturber moves into the supersonic regime. However, orbits with eccentricities $e\gtrsim 0.8$ below the sound barrier experience a slight increase in eccentricity as they loose orbital energy. Possible extensions to our analytical approach are also discussed.
\end{abstract}



\section{Introduction}

Dynamical friction (DF) arises from the gravitational backreaction induced by the motion of a "perturber" (a compact object, a satellite galaxy etc.) in a discrete or continuous medium (of stars, gas, dark matter etc.). 
It is ubiquitous in cosmic structure formation, with applications ranging from the dynamical evolution of planetisimals, stars and their remnants on sub-parsec scales to the merging of galaxies on mega-parsec scales \cite[see e.g.][]{tremaine/etal:1975,binney/tremaine:1987,kauffmann/etal:1993,somerville/primack:1999,cole/etal:2000,goldreich/etal:2004,croton/etal:2006,boylankolchin/etal:2008,Kaur2022}. In a pioneering paper, \cite{chandrasekhar:1943} derived an expression for the DF force produced by a point-like perturber moving in linear motion in a collisionless medium. Chandrasekhar's result has been widely applied and extended to other astrophysical systems. includes gaseous media \citep{dokuchaev:1964,ruderman/spiegel:1971,rephaeli/salpeter:1980,just/kegel:1990,ostriker:1999,sanchez/brandenburg:2001,kim/etal:2007,lee/stahler:2011,vicente/etal:2019,ssalcedo:2019,desjacques/etal:2022,macleod/etal:2022} and, more recently, backgrounds of axion dark matter \citep{hui/etal:2017,baror/etal:2019,chavanis:2021,traykova/clough/etal:2021,buehler/desjacques:2023,foote/etal:2023,tomaselli/etal:2023,traykova/vicente/etal:2023}. 

Most theoretical studies thus far have assumed that the perturber moves in linear motion. Exact solutions such as e.g. \cite{ostriker:1999}'s are routinely applied to model the impact of DF on proto-planetary systems or on the dynamics of compact stellar binaries \citep[see for instance][]{Iben_1993,grishin/perets:2015,staff2016, grishin/perets:2016,MacLeod_2017,Antoni2019,Ginat/Glanz2020,De2020,Everson_2020,Rozner2022}. However, it would be very desirable to extend the scope and validity of the theoretical results to generic bound (eccentric) orbits. Several pieces of work have investigated the DF experienced by circularly-moving perturbers using a variety of analytical and numerical methods for both collisionless and collisional media \citep[see for instance][]{tremaine/weinberg:1984,sanchez/brandenburg:2001,kim/etal:2007,kim/etal:2008,kaur/sridhar:2018,ssalcedo:2019,banik/vandenbosch:2021,desjacques/etal:2022}. Using linear response theory, \cite{desjacques/etal:2022} developed an analytical approach to compute the DF for a circular motion in a gaseous medium. The salient differences with the corresponding linear motion formula are the absence of a far-field, logarithmic divergence and the appearance of a radial (i.e. perpendicular) component in the DF force. Like the linear-motion result however, the steady-state circular DF peaks for a Mach number $\mach\simeq 1$. Therefore, if the steady-state approximation to DF holds, the orbit of a perturber moving on a bound eccentric trajectory should rapidly circularize as the perturber looses orbital energy and increasingly moves at supersonic speed. 

To investigate this issue further, we build on the approach of \cite{desjacques/etal:2022} to explore Dynamical Friction when the orbital eccentricity is significant. 
The paper is organized as follows. Section \S\ref{sec:theory} summarizes our computation of the friction coefficient for a generic elliptic orbit; Section \S\ref{sec:validation} shows that our analytical approximation is valid for a range of Mach numbers $\mach\sim 1$; In Section \S\ref{sec:OrbEvol} we apply our results to eccentric orbits to study their evolution under the effect of DF; We summarize our results and conclude in Section \S\ref{sec:conc}.

\section{From circular to elliptic orbits}

\label{sec:theory}


\subsection{General relations}

Following \cite{ostriker:1999,desjacques/etal:2022}, the DF force in Newtonian gravity can be generally expressed as
\begin{equation}
    \label{eq:Newton}
    \fdf(t)=G M \bar\rho_g \int\!\mathrm d^3 u \frac{\vu}{u^3} \alpha(\vu ,t)
\end{equation}
where $\bar\rho_g$ is the density of the unperturbed (uniform) gaseous medium, $\vu=\vr-\vr_p(t)$ is the separation vector relative to the current position $\vr_p(t)$ of the perturber, and $\alpha(\vr,t)$ is the fractional gas density perturbation. In the linear response theory considered here, $\alpha(\vr,t)$ solves the driven, linearized sound wave equation
\begin{equation}
\label{eq:drivensoundwaveeq}
\frac{\partial^2\alpha}{\partial t^2} - c_s^2 \nabla^2\alpha = 4\pi GM\,h(t)\,\delta^D(\vr-\vr_p(t)) \;.
\end{equation}
Here, $c_s$ is the speed of sound, whereas $h(t)$ is 1 if the perturber is active and zero otherwise. 
Transforming to Fourier space and applying Green's method, we can solve for the overdensity and, thereby, express the DF force as 
\begin{align}
    \label{eq:generalDF}
    \fdf(t) &= \big(4\pi GM\big)^2 \bar\rho_g 
    \int_\omega \int_{-\infty}^{+\infty}\!dt'\int_{\vk}\,h(t')\,
    \frac{i\vk}{k^2} \\
    &\qquad\times    \frac{e^{i\vk\cdot(\vr_p(t)-\vr_p(t'))-i\omega(t-t')}}{c_s^2k^2-(\omega+i\epsilon)^2} \nonumber\;,
\end{align}
after taking advantage of the Fourier transform $\int\! \mathrm d^3 u \frac{\vu}{u^3}e^{i \vk\cdot \vu}=4\pi\frac{i\vk}{k^2} $ of the Coulomb potential. 
We have also defined $\intw =\frac{1}{2\pi}\int_{-\infty}^{\infty} \mathrm{d}\omega$, and 
\begin{equation}
\int_{\vk}=\frac{1}{(2\pi)^3}\int_0^{2\pi}\mathrm{d}\varphi_k \int_{-1}^1 \mathrm{d} \cos(\vartheta_k)\int_0^\infty \mathrm{d}k\ k^2
\end{equation}
in spherical coordinates for which $\vk=(k,\varphi_k,\vartheta_k)$.
Eq.~(\ref{eq:generalDF}) is still completely general as far as the orbital motion $\vr_p(t)$ is concerned.

\begin{figure*}
    \centering
    \hspace*{-1.3cm}
    \includegraphics[width=0.8\textwidth,height=50em]{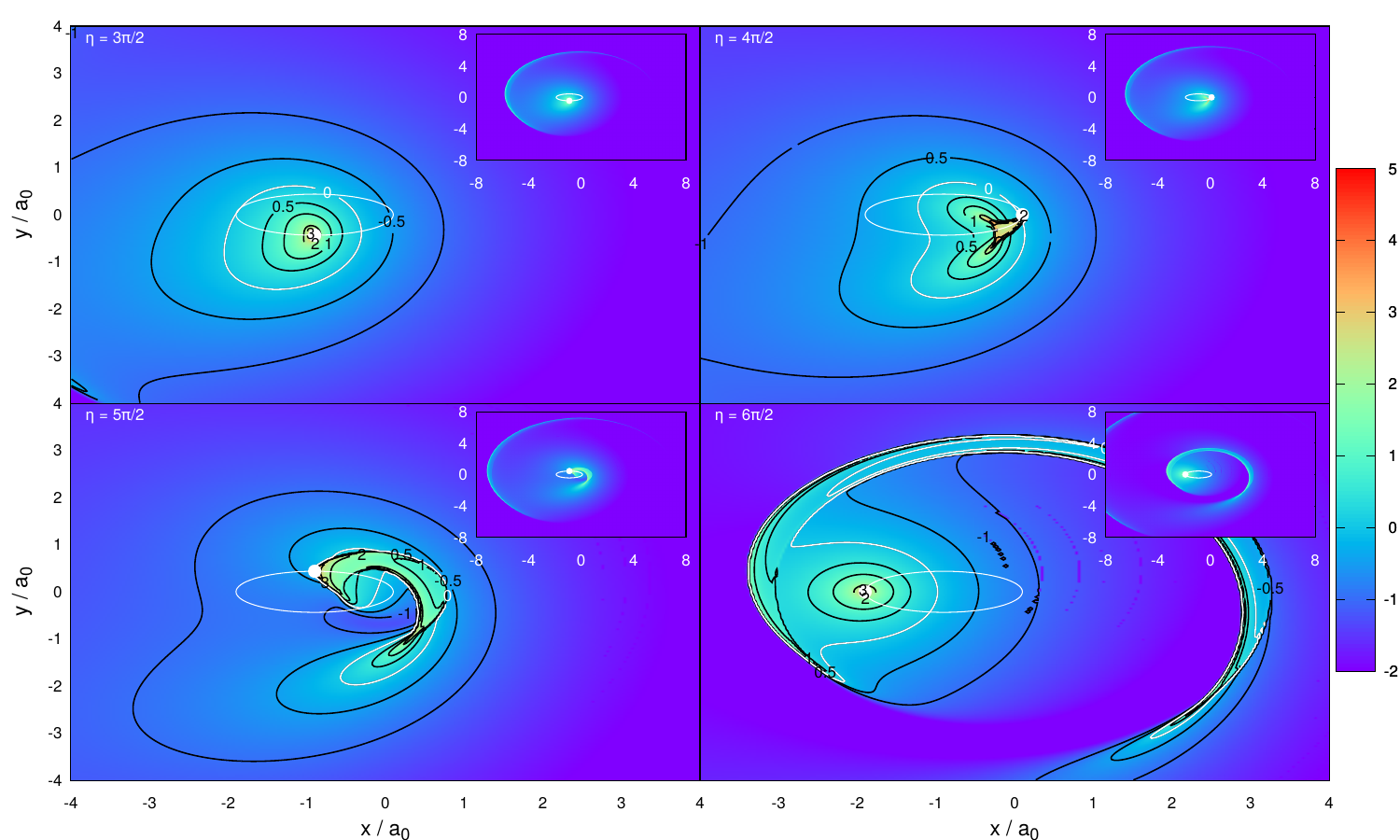}
    \caption{Evolution of the gas overdensity $\alpha(\vr,\eta)$ computed from Eq.~(\ref{eq:alphaElip}) for a perturber with characteristic Mach Number $\mach_a=0.9$ and eccentricity $e=0.9$. The semi-major axis factorizes out and is thus left unspecified. Snapshots of $\log(\alpha)$ (represented by the color scale) are shown in the orbital plane at four successive times given by $\eta=\frac{3\pi}{2}$ (top left panel), $2\pi$, $\frac{5\pi}{2}$ and $3\pi$ (bottom right panel). The orbit and the position of the perturber are indicated by a curve and a white circle, respectively. We omit the first half rotation because the wake has not fully developed by that time and is thus not very informative. In each panel, a zoomed-out inset shows the evolution of the far-field density wake as it moves outward the orbit.}
    \label{fig:alphaElip}
\end{figure*}

\subsection{DF for eccentric orbits}

Since we are interested in a perturber on a bound eccentric orbit, it is convenient to parameterize the latter with the eccentric anomaly $\eta$. Assuming that the motion takes place in the $x-y$ plane, we have
\begin{align}
    r(\eta) &= a \big(1-e\cos\eta\big) \\
    \cos\vartheta(\eta) &= \frac{\cos\eta-e}{1-e\cos\eta}\;,\quad \sin\vartheta(\eta) = \frac{\sqrt{1-e^2}\sin\eta}{1-e\cos\eta} \nonumber \\
    t(\eta) &=\omega^{-1}\big(\eta-e\sin(\eta)\big) \nonumber
\end{align}
where $a$, $e$ and $\vartheta$ are the semi-major axis, eccentricity and true eccentric anomaly respectively. For a perturber orbiting a (massive) companion (located at the origin of coordinates) counterclockwise, the position vector of its eccentric orbit is 
\begin{align}
    \label{eq:rp}
    \vr_p(\eta) &= a \big(\cos\eta-e\big)\, \xvh + a\sqrt{1-e^2}\sin\eta\, \yvh \\
    &= \vr_c(\eta) - a e\, \xvh - a \big(1-\sqrt{1-e^2}\big) \sin\eta\, \yvh \nonumber \;,
\end{align}
in which
\begin{equation}
    \vr_c(\eta) \equiv a \cos\eta\, \xvh + a\sin\eta\,\yvh
\end{equation}
delineates a circular orbit ($e=0$) with identical semi-major axis. A non-zero eccentricity thus perturbs the circular orbit in two ways: it changes i) the physical shape of the orbit (from a circle to an ellipse) and ii) the time lapse along the orbit. As we will see shortly, the second effect dominates across a range of Mach number for which it is possible to derive an accurate prediction for the DF force.

Parameterizing the orbit with the mean anomaly, the DF force can be expressed as
\begin{align}
    \label{eq:FDF1}
    \fdf(\eta) &=\big(4\pi GM\big)^2 \bar\rho_g \int_\omega \int_{-\infty}^{+\infty}\!d\eta'\,h\big(t(
    \eta')\big)\,\Omega^{-1}\big(1-e\cos\eta'\big) \nonumber \\
    &\qquad\times    \int_{\vk}\,\frac{i\vk}{k^2}\,
    \frac{e^{i\vk\cdot(\vr_p(\eta)-\vr_p(\eta')-i\omega(t(\eta)-t(\eta'))}}{c_s^2k^2-(\omega+i\epsilon)^2} \;.
\end{align}
The Rayleigh decomposition of $e^{i\vk\cdot(\vr_p(\eta)-\vr_p(\eta')}$ is particularly powerful for the circular case \citep[see][]{desjacques/etal:2022} since $\fdf(\eta)$ can then be conveniently expanded on the (spherical) helicity basis $\{\zvh,\evh_+,\evh_-\}$ with $\evh_\pm=\frac{1}{\sqrt{2}}( i\yvh\mp\xvh)$, 
\begin{equation}
    \fdf(\eta)=F^{(0)}(\eta)\, \zvh + F^{(+)}(\eta)\,\evh_+ + F^{(-)}(\eta)\,\evh_- \;.
\end{equation} 
This decomposition can also be used in the eccentric case, although the variation of the orbital radius $r(\eta)$ makes the calculation tedious. On substituting 
\begin{equation}
\vk = \sqrt{\frac{4\pi}{3}} k \left(Y_1^0(\kvh)\,\zvh+Y_1^{+1}(\kvh)\,\evh_+ + Y_1^{-1}(\kvh)\,\evh_-\right)
\end{equation}
into Eq.~(\ref{eq:FDF1}) and performing the Gaunt integral, we arrive at
\begin{align}
     F^{(+1)}\!(\eta) &= 4\pi \left(\frac{GM}{\Omega a}\right)^2\, \bar\rho_g\, \frac{e^{i\eta}}{\sqrt{2}}\, I\big(\mach_a,e,\eta\big) \\
    F^{(-1)}\!(\eta) &= - F^{(+1)*}\!(\eta) \nonumber \\
    F^{(0)}\!(\eta)&=0 \bigg. \nonumber \;.
\end{align}
Here,
\begin{equation}
    \label{eq:Mach}
    \mach_a = \frac{\Omega a}{c_s} = \frac{1}{c_s}\sqrt{\frac{GM_\bullet}{a}}
\end{equation}
is a characteristic Mach number~\footnote{It is the Mach number of a perturber moving on a circular orbit of radius $a$.} and $M_\bullet\gg M$ is the mass of the companion.
The (complex) friction coefficient $I(\mathcal{M}(a),e,\eta)$ encodes the dependence of the DF force on the nature of the medium and the value of the orbital elements. Appendix \S\ref{app:friction} outlines an approximation to the steady-state friction coefficient, which captures timing variation in the orbit (i.e. $t(\eta)$) relative to the circular case but neglect the change in the orbit radius (i.e. $r(\eta)$)
The final expression of $I(\mathcal{M}(a),e,\eta)$ is given by the multipole expansion (\ref{eq:Im}) and (\ref{eq:SMetae}).
Appendix \S\ref{app:friction} also demonstrates that this expansion has a short distance logarithmic divergence, which is regulated by truncating the series at some maximum multipole $\ell_\text{max}$.

Projecting the force onto the instantaneous radial and tangential directions $\evh_r(\eta)=\cos\vartheta(\eta)\,\xvh+\sin\vartheta(\eta)\,\yvh$ and $\evh_\vartheta(\eta)=-\sin\vartheta(\eta)\,\xvh+\cos\vartheta(\eta)\,\yvh$, and using the relation
\begin{equation}
    \evh_{\pm} = \frac{e^{\mp i\vartheta(\eta)}}{\sqrt{2}}\big(\mp \evh_r(\eta) + i \evh_\vartheta(\eta)\big) \;,
\end{equation}
we eventually obtain
\begin{equation}
    \label{eq:FDF_final}
    \fdf(\eta) = F_r(\eta) \evh_r(\eta) + F_\vartheta(\eta) \evh_\vartheta(\eta)\;,
\end{equation}
where 
\begin{align}
    \label{eq:F_rF_v}
    F_r(\eta) &= -4\pi \left(\frac{GM}{\Omega a}\right)^2\, \bar\rho_g\,\Re\!\left(e^{i(\eta-\vartheta(\eta))}\,I\big(\mach_a,e,\eta\big)\right) \\
    F_\vartheta(\eta) &= -4\pi \left(\frac{GM}{\Omega a}\right)^2\, \bar\rho_g\, \Im\!\left(e^{i(\eta-\vartheta(\eta))}\,I\big(\mach_a,e,\eta\big)\right) \nonumber
\end{align}
are the radial and azimuthal components of the DF force along the trajectory of the perturber. Note that the instantaneous, radial unit vector $\evh_r(\eta)$ is directed outward, while the azimuthal unit vector $\evh_\vartheta(\eta)$ points in the direction of the (counterclockwise) motion.

\section{Validation with simulations}

\label{sec:validation}

To validate our approximation, we compute the DF force after solving the driven sound wave equation (\ref{eq:drivensoundwaveeq}) on a 3-dimensional grid. 

Using the retarded Green's function, we calculate the overdensity $\alpha$ on a regular, $512^3$ cubical mesh of length $16a$ centered on the massive companion, i.e.
\begin{align}
    \alpha(\vr_i,\eta) &=\frac{G M}{c_s^2}\int_0^{\eta} \mathrm{d}\eta'(1-e \cos(\eta')) \frac{\delta^{D}(\eta'-\frac{1}{c_s}|\vr_i-\vr_p(\eta')|}{|\vr_i-\vr_p(\eta')|}\nonumber\\
    &\approx\frac{G M}{\sqrt{2\pi}\sigma c_s^2}\int_0^{\eta} \mathrm{d}\eta'(1-e \cos(\eta')) \frac{e^{-\frac{(\eta'-\frac{1}{c_s}|\vr_i-\vr_p(\eta')|)^2}{2\sigma^2}}}{|\vr_i-\vr_p(\eta')|}\;.
    \label{eq:alphaElip}
\end{align}
where $\vr_i$ are discretized grid coordinates, $\vr_p(\eta)$ given by Eq.~(\ref{eq:rp}) is the position of the perturber and $\eta$ plays the role of the clock.
The second equality follows from approximating the Dirac-delta distribution with a Gaussian of width of $\sigma=0.01a$. The simulations assume absorbing boundary conditions at the outer edge of the grid and no accretion on the perturber. They implement the finite time perturbation such that $h(\eta)=1$ for $\eta>0$ and zero otherwise. 

Fig.~\ref{fig:alphaElip} displays the evolution of the gas fractional density fluctuation $\alpha(\vr,t)$ in the orbital plane for an elliptic orbit with $(\mach_a,e)=(0.9,\ 0.9)$. Snapshots are shown at four different times corresponding to eccentric anomalies $\eta=3\pi/2$, $2\pi$, $5\pi/2$ and $3\pi$ as indicated in the figure. The instantaneous Mach number 
\begin{equation}
    \label{eq:Machinstant}
    \mach(\eta) = \mach_a \sqrt{\frac{1+e\cos\eta}{1-e\cos\eta}}
\end{equation}
is $\mach(\eta)= 3.9$ (resp. 0.2) at pericenter (resp. apocenter).
At $\eta=\frac{3\pi}{2}$ the near-field density wake (in the vicinity of the perturber) is nearly circular, leading to a DF force which is close to zero. As the perturber passes through the pericenter, the near-field wake becomes asymmetrical and elongated while the supersonic motion of the perturber produces a Mach cone. All this causes the DF force to rise. The Mach cone lasts until apocenter passage, where the motion becomes subsonic again while the trailing density wake detaches from the perturber and propagates outwards as a spiral shock wave. The fairly symmetric distribution of the near- and far-field density wakes at apocenter minimizes the DF force. In the zoomed-out insets of Fig.~\ref{fig:alphaElip}, the spiral shock wave which detached at the first apocenter passage ($\eta=\pi$) can be seen propagating outwards. Note also that the wake density always exceeds the average density, i.e. $\alpha(\vr,\eta)\geq 0$ everywhere. This arises from the fact that the Green's function is positive definite ($\propto 1/r$) and the perturber is an overdense perturbation.

\begin{figure*}
    \centering
    \includegraphics[width=0.8\textwidth,height=50em]{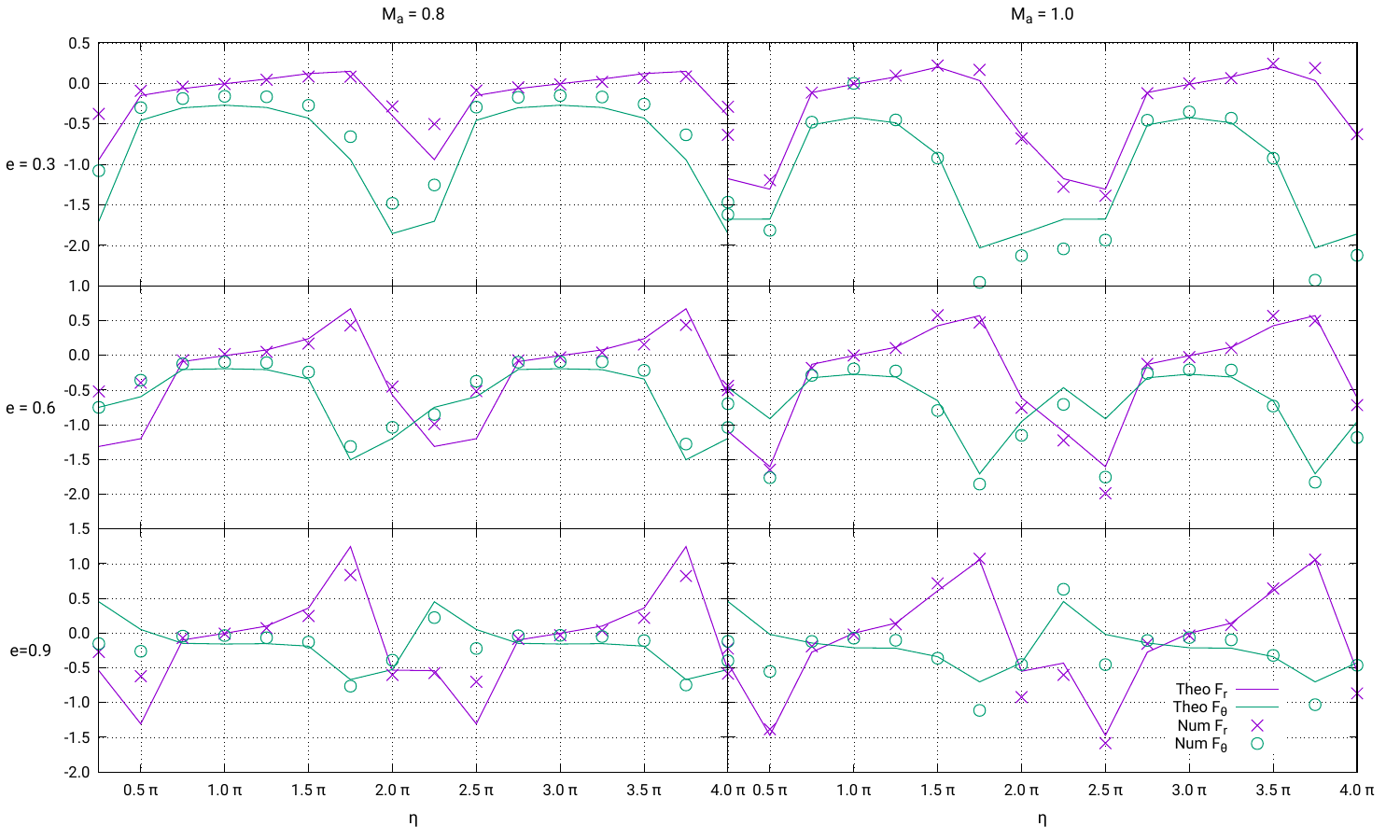}
    \caption{Comparison between the numerical simulation and the theoretical prediction of $F_r$ and $F_\vartheta$ (in units $4 \pi \bar \rho_g \mach_a^2 \left(\frac{G M}{\Omega a}\right)^2$) for all combination of $e = \{0.3,\ 0.6,\ 0.9\}$ (rows top to bottom) and $\mach = \{0.8,\ 1.0\}$ (columns left to right). The simulation results implement the finite time perturbation and are shown for the first two rotations of the perturber. The theoretical, steady-state prediction matches well the overall behaviour of the numerical data for all parameter combination, although it is not able to always reproduce the exact values. These discrepancies grow with eccentricity and are most pronounced around pericenter passage.}
    \label{fig:elip_num_theo}
\end{figure*}

Using Eq.~\ref{eq:Newton}, we calculate the DF force acting on the perturber for each 3-dimensional snapshot of the overdensity field $\alpha(\vr,t)$. First, as a consistency check, we tested our simulation setup for the circular case $e=0$ to ensure that the size of the box and the resolution are sufficient enough to properly capture the DF force. Our simulation setup successfully recovers the  analytical results of \cite{desjacques/etal:2022} when the largest multipole $\ell_\text{max}\sim \pi/(\Delta/a)$ is matched to the mesh resolution $\Delta= a/32$. Next, we produced a suite of "simulations" for the parameter choices $e\in[0.3,\ 0.6,\ 0.9]$ and $\mach_a\in[0.8,\ 1.0]$. A comparison between the "simulated" DF force and the analytical approximation based on equations (\ref{eq:Im}) and (\ref{eq:SMetae}) is presented in Fig.~\ref{fig:elip_num_theo}. The latter assumes steady-state, and only takes into account the dependence of $t(\eta)$ on eccentricity (i.e. $\vr_p(\eta)\simeq\vr_c(\eta)$ as discussed in Appendix \S\ref{app:friction}).

For the finite time perturbation implemented by the simulations, the asymmetry of the perturber's trajectory suggests that, unlike the circular case for which steady-state is achieved exactly after one sound-crossing time $t_\text{sc}=2a/c_s$ of the system \citep[see][]{desjacques/etal:2022}, convergence to steady-state may occur on a different timescale when $e>0$. Notwithstanding, our theoretical predictions appear to reproduce the numerical results reasonably well for the parameter combinations considered here, although discrepancies can be seen at large eccentricities especially around pericenter passage. In general our solution tends to overestimate $F_\vartheta$ while it underestimates $F_r$. For Mach numbers outside the range $[0.75,1.05]$, we have found that our analytical approximation to DF is a poor match to the numerical simulation regardless the eccentricity. 

Fig.~\ref{fig:elip_num_theo} also shows that $F_r$ reaches a (positive) maximum (the radial force is thus directed outward) in the time interval $\frac{3\pi}{2}\lesssim \eta\lesssim 2\pi$, which coincides with the minimum of $F_\vartheta$. $F_r$ changes then abruptly at pericenter passage and reaches a minimum for $\eta\simeq \frac{\pi}{2}$, which is somewhat delayed relative to the maximum of $F_\vartheta$. The latter turns out to be positive for $e=0.9$ so that the azimuthal component is directed along the direction of motion (and thus increases the kinetic energy of the perturber). Note that these extrema occur along the orbit approximately when the instantaneous Mach number of the perturber becomes larger or smaller than its orbit averaged value $\mach_a$ (i.e. $\eta=\pi/2$ and $3\pi/2$.)

\begin{figure}
    \includegraphics[width=0.5\textwidth]{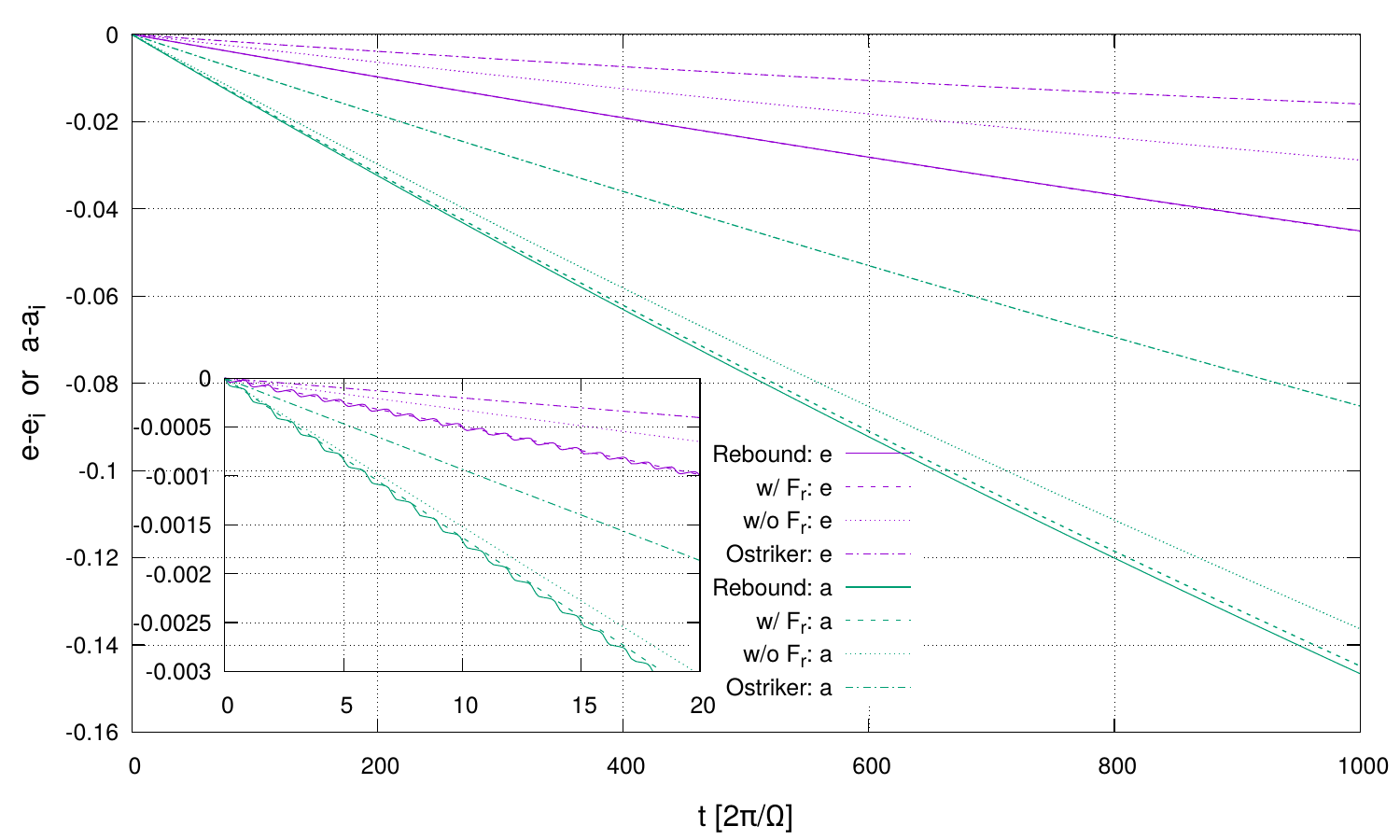}
    \caption{Evolution of the semi-major axis and eccentricity across 1000 orbits assuming an initial eccentricity $e_i=0.3$ and Mach number $\mach_i=0.9$, a binary mass ratio $q=10^{-3}$ and a uniform gas density $\tilde \rho_g=10^{-3}$. The solid line was obtaining by evolving the orbit with the N-body integrator {\small REBOUND} with the instantaneous components $F_r(\eta)$ and $F_\vartheta(\eta)$ of the DF force given by Eq.~(\ref{eq:FDF_final}). The dashed line represents the solution to the coupled ODEs~(\ref{eq:orbAvg2}) obtained from the orbit-averaged frictions $\overline{I_a}$ and $\overline{I_e}$; the dashed-dotted line shows the effect of ignoring $F_r$ in the calculation of $\overline{I_a}$ and $\overline{I_e}$; the dotted line shows the effect of using the linear motion solution of \protect\cite{ostriker:1999} for the computation of $\protect \overline{I_a}$ and $\overline{I_e}$. The zoomed-in inset focuses on the first 20 rotations.}
    \label{fig:rebound}
\end{figure}

\section{Long-term orbital evolution}
\label{sec:OrbEvol}

In spite of its limited range of validity, our approximation to the eccentric DF force can be used to calculate the evolution of orbital eccentricity as the perturber crosses the sound barrier.

It is convenient to use dimensionless units in order to calculate the evolution of the orbital parameters. For this purpose, we introduce a characteristic semi-major axis $a_0$ and frequency $\Omega_0$, which are related through Kepler's third law $\Omega_0=(GM_\bullet)^{1/2}a_0^{-3/2}$. They define the dimensionless variables 
\begin{equation}
    \tilde a(t) = \frac{a(t)}{a_0}\;,\qquad \tilde t=t \Omega_0 \;,\qquad \tilde \Omega(t) = \tilde a(t)^{-3/2} \;,
\end{equation}
which we use in the numerical implementation below. 

\subsection{Evolution of the orbital parameters}

The change in the orbital parameters are governed by \citep{burns:1976,dermott/murray:1999}
\begin{align}
    \frac{\mathrm d\tilde a}{\mathrm d\tilde t}&=2\sqrt{\frac{\tilde a^3}{(1-e^2)}}\, q^{-1}\left[ \tilde F_r e \sin\vartheta+ \tilde F_\vartheta \big(1+e \cos\vartheta\big) \right]\\
    \frac{\mathrm de}{\mathrm d \tilde t}&=\sqrt{\tilde a(1-e^2)}\, q^{-1}\left[\tilde F_r \sin\vartheta+\tilde F_\vartheta\big(\cos\vartheta+\cos\eta\big) \right] \nonumber \;.
\end{align}
Here, $q=M/M_\bullet\ll 1$ is ratio of the perturber's to the massive companion's mass and $\tilde F_{r,\vartheta}=\frac{F_{r,\vartheta}}{M_\bullet a_0 \Omega_0^2}$ are the components of a normalized DF force.
Since the latter are given by Eq.~(\ref{eq:F_rF_v}), averaging the rate of change of the orbital elements over one period gives 
\begin{align}
    \label{eq:orbAvg2}
    \left\langle \frac{\mathrm d\tilde a}{\mathrm d\tilde t}\right\rangle &=-4\, \tilde \rho_g\, q\,\tilde a^{5/2}\,(1-e^2)^{-1/2}\, \overline{I_a}(\tilde a, e)\\
  \left\langle \frac{\mathrm de}{\mathrm d \tilde t}\right\rangle &=-2\, \tilde \rho_g\, q\,\tilde a ^{3/2}\, (1-e^2)^{1/2}\, \overline{I_e}(\tilde a,e) \nonumber \;,
\end{align}
where $\tilde \rho_g=\bar \rho_g \frac{a_0^3}{M_\bullet}$ is a normalized gas density and we have defined the orbit averaged friction coefficients
\begin{align}
    \label{eq:orbAvgIa}
    \overline{I_a}(\tilde a,e)&=\frac{1}{2\pi}\int_0^{2\pi} \!\mathrm d \eta\ (1-e \cos\eta) \\&\qquad \cdot \bigg[  \Re\left(e^{i (\eta-\vartheta)}\, I\big(\mach_{\tilde a},e,\eta\big)\right )\, e \sin\vartheta \nonumber \\
    &\qquad + \Im\left(e^{i(\eta-\vartheta)}\,I\big(\mach_{\tilde a},e,\eta\big)\right)(1+e \cos\vartheta)\bigg] \nonumber 
\end{align}
and
\begin{align}
    \label{eq:orbAvgIe}
    \overline{I_e}(\tilde a,e)&=\frac{1}{2\pi}\int_0^{2\pi} \!\mathrm d \eta\ (1-e \cos\eta) \\ &\qquad\cdot\bigg[\Re\left(e^{i(\eta-\vartheta)}\, I\big(\mach_{\tilde a},e,\eta\big)\right)\, \sin\vartheta \nonumber \\
    &\qquad +\Im\left(e^{i(\eta-\vartheta)}\, I\big(\mach_{\tilde a},e,\eta\big)\right)(\cos\vartheta+\cos\eta)\bigg] \nonumber \;,
\end{align} 
with $\mach_{\tilde a}=\mach_{a(\tilde a)}=\tilde\Omega \tilde a (\Omega_0 a_0/ c_s)$ (\ref{eq:Mach}).
Since these orbit averaged quantities must be evaluated numerically, we found prudent to check our results with the high-precision N-body integrator {\small REBOUND} \citep{REBOUND}. 

For this purpose, we set it up with one central mass and a perturber with $q=10^{-3}$. The initial ($t=0$) position and velocity match an unperturbed, elliptic orbit with eccentricity $e=0.3$ and orbit averaged Mach number $\mach_{\tilde a}=0.9$. For $t>0$, we apply, in addition to the gravitational pull of the central mass, the DF force the perturber would experience if it were moving in an uniform gaseous medium of density $\tilde \rho_g=10^{-3}$. The smallness of the product $q\tilde\rho_g=10^{-6}$ ensures that the orbital parameters $e$ and $\tilde a$ vary on a timescale significantly longer than the dynamical time, so that steady-state Dynamical Friction holds. Therefore, we shall assume the steady-state approximation to the DF force given in Appendix \S\ref{app:friction} throughout. The component of the DF force are calculated according to Eq.~(\ref{eq:F_rF_v}) using the instantaneous eccentricity and semi-major axis provided by {\small REBOUND}. 

The results of this simulation are displayed in Fig.~\ref{fig:rebound} as the solid curves. 
These are compared to the solution to the coupled ODEs~Eq.~(\ref{eq:orbAvg2}) with $\overline{I_a}$ and $\overline{I_e}$ calculated i) following equations (\ref{eq:orbAvgIa}) and (\ref{eq:orbAvgIe}) (dashed line) (ii) ignoring the imaginary part (dotted line) and (iii) using the (purely complex) friction coefficient $I=I(\mach_{\tilde a})$ derived by \cite{ostriker:1999} in the linear motion case (dashed-dotted line). Unsurprisingly, case (i) matches best the instantaneous evolution given by {\small REBOUND}: the eccentric evolution is accurately reproduced, while the evolution of the semi-major axis deviates only by $\approx 1.5\%$ after 1000 orbits. Case (ii) demonstrates that discarding only the real part or, equivalently, the radial component $F_r$ already leads to a noticeable deviation in the evolution of the orbital parameters. The discrepancy is even larger for case (iii), for which the real part is zero while the imaginary part is computed from the linear-motion solution of \cite{ostriker:1999}.

\begin{figure*}
    \centering
    \includegraphics[width=0.9\textwidth]{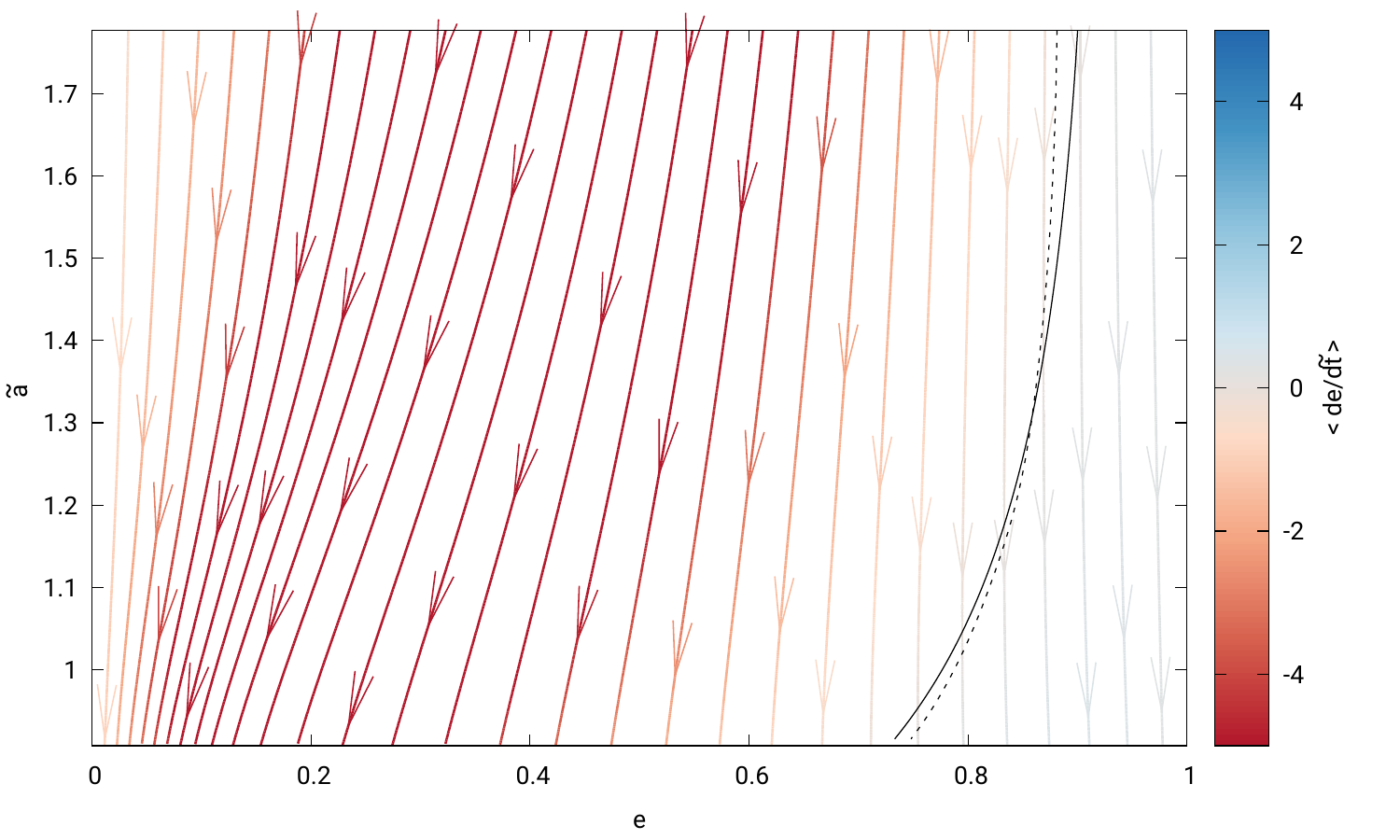}
    \caption{Integral curves in the $(e,\tilde a)$ plane obtained by solving the system of ODEs \ref{eq:orbAvg2}. We have assumed $c_s/a_0\Omega_0=q\tilde \rho_g=1.0$, which implies $\mach_{\tilde a}=\tilde a^{-1/2}$. The range of $\tilde a$ is chosen such that our analytical approximation is viable. Colors represent the rate $\langle de/d\tilde t\rangle$ of eccentricity change. The overall magnitude is arbitrary since $\langle de/ d\tilde t\rangle\propto q\tilde \rho_g$. However, the locus where $\langle de/ d\tilde t\rangle=0$ shown as the black curve is robust to the choice of $q\tilde\rho_g$. On the left of it, orbits circularize while, on the right of it, their eccentricity increases. The dotted curve is an analytical estimate of this boundary based on the change of orbital energy and angular momentum (see text for details).}
    \label{fig:odeflow}
\end{figure*} 

\subsection{Eccentric evolution for Mach numbers $\mach\sim 1$}

Fig.~\ref{fig:odeflow} shows the integral curves defined by the flow equations~(\ref{eq:orbAvg2}) assuming an initial eccentricity in the range $0<e_i<1$ but a unique, initial semi-major axis $\tilde a_i\approx 1.8$ corresponding to a Mach number $\mach_{\tilde a}=0.75$. Furthermore, since the vector flow is independent of the product $\tilde \rho_g q$ (which can be absorbed into a redefinition of the time coordinate), we have set $\tilde\rho_g q = 1$ without loss of generality. Since the perturber loses energy regardless of the choice of $e_i$ and $\tilde a_i$ (DF transfers orbital energy to the density wake), the orbit always shrinks to smaller semi-major axes. As a result, the characteristic Mach number eventually exceeds the upper bound above which our approximation ceases to be accurate. This occurs when $\tilde a(t)\approx0.9$, at which point we stop the computation of the integral curves.

The eccentric evolution is sensitive to the choice of $e_i$. For $e_i\lesssim 0.7$, the orbit tends to circularize by the time $\mach_a$ exceeds unity, with an effect strongest in the range $0.2\lesssim e_i\lesssim 0.4$. For $e_i\gtrsim 0.7$, the orbit becomes more eccentric as can be seen from the solid (black) curve, which marks the locus for which ${\rm d}e/{\rm d}t=0$. Fig.~\ref{fig:elip_num_theo} suggests a simple, intuitive explanation: near pericenter passage, the azimuthal component $F_\vartheta$ can be positive at high eccentricities. This increases the kinetic energy (i.e. the orbital energy) of the perturber and, thereby, the distance of the apocenter. As a result, the orbit becomes more elliptic. The converse is true at low eccentricities: $F_\vartheta$ is negative and thus slows down the perturber near pericenter passage, which tends to circularize the orbit.

In order to quantify this further, we follow the analytical argument of \cite{macleod/etal:2022} and introduce the specific angular momentum $h=\sqrt{G M_\bullet a (1-e^2)}$ and orbital energy $\varepsilon=-\frac{G M_\bullet}{2 a}$. This allows us to express the eccentricity as
\begin{equation}
    e^2=1+\frac{2 \varepsilon h^2}{(G M_\bullet)^2}\;.
\end{equation}
A body subject to dynamical friction experiences a change of energy
\begin{equation}
    \Delta \varepsilon =\frac{\mathbf v_p(\eta) \cdot \fdf}{M}\Delta t\;
\end{equation}
where the velocity is given by 
\begin{equation}
\vv_p(\eta)=\frac{\Omega a}{1-e \cos\eta} \big(-\sin\eta\, \xvh + \sqrt{1-e^2}\cos\eta\, \yvh\big) 
\end{equation}
Furthermore, DF generates a torque which changes the angular momentum by
\begin{equation}
    \Delta h=\frac{\vr_p(\eta) \times \fdf}{M}\Delta t\;.
\end{equation}
Therefore, DF changes $u=e^2-1$ (which is a proxy for the eccentricity) by
\begin{equation}
    \frac{\Delta u}{u}\approx \frac{\Delta \varepsilon}{\varepsilon}+\frac{2\Delta h}{h}\;.
\end{equation}
Rather than integrating over a whole orbit, the change of $\frac{\Delta u}{u}$ can be estimated from the empirical observation that $\Delta u/u$ reaches a positive maximum at $\eta=\pi/2$ and negative minimum at $\eta=\pi$. In other words, the loss of eccentricity is maximum at $\eta=\pi/2$, while the gain of eccentricity is largest at $\eta=\pi$. We thus write
\begin{equation}
    \frac{\Delta u}{u}\bigg\lvert_{\rm tot}\approx\frac{\Delta u}{u}\bigg\lvert_{\eta=\pi}+\frac{\Delta u}{u}\bigg\lvert_{\eta=0.5\pi} \;.
    \label{eq:utot}
\end{equation}
Approximating the time interval during which the DF force acts on the body as 
\begin{equation}
\Delta t\approx \frac{a}{|\mathbf v_p|}=\Omega^{-1} \sqrt{\frac{1-e \cos \eta}{1+e \cos\eta}}
\end{equation}
and using our analytical solution to the DF force provides an estimate for $\Delta\epsilon$ and $\Delta h$ when the gain/loss of eccentricity is maximum and, thereby, an estimate for $\Delta u/u|_{\rm tot}$ as given by Eq.~\ref{eq:utot}. Setting $\Delta u/u|_{\rm tot}=0$ gives the locus shown as the dotted line in Fig.~\ref{fig:odeflow}, for which the gain and loss of eccentricity balance each other, i.e. ${\rm d}e/{\rm d}t=0$. This prediction is in good agreement with that inferred from the computation of the integral curves (solid black curve). 

Summarizing, most trajectories will tend to circularize as the sound barrier is crossed. However, orbits with $e\gtrsim 0.8$ at characteristic Mach number $\mach_a\sim 0.8$ experience a (slight) increase in eccentricity while the perturber looses orbital energy and moves into the supersonic regime. 

\section{Discussion and Conclusions}

We have investigated the effect of dynamical friction (DF) for a perturber moving on a bound eccentric orbit in a gaseous medium. We have extended the multipole approach of \cite{desjacques/etal:2022} to capture timing variations relative to the circular case through a perturbative expansion in the orbital eccentricity (the "small" parameter) $e$. However, we have not succeeded in capturing the physical deformation of the orbit (which breaks the planar symmetry) and have thus neglected it. 

We have validated our analytical (steady-state) approximation based on timing variations with measurements of the DF force extracted from 3-dimensional simulations of the gas density response. 
We have found good agreement for characteristic Mach numbers $\mach_a=\Omega a/c_s$ ($a$ is the ellipse semi-major axis) in the range $0.75\lesssim \mach_a\lesssim 1.05$, even for eccentricities as large as $e=0.9$. The reason why the timing variation dominates in this range of characteristic Mach number has remained elusive. The observed agreement indicates also that the finite time perturbation implemented by the 
3-dimensional simulations approaches steady-state on a dynamical timescale, that is, the sound-crossing time of the system as in the circular case \cite[see][]{kim/etal:2007,desjacques/etal:2022}. 
Furthermore, snapshots of the gas density response show that, at high eccentricities, the trailing density wake induced by the perturber is a series of concentric, incomplete ring-like patterns produced in "bursts" around pericenter passage.

\label{sec:conc}
\begin{figure}
    \centering
    \includegraphics[width=.5\textwidth,height=7cm]{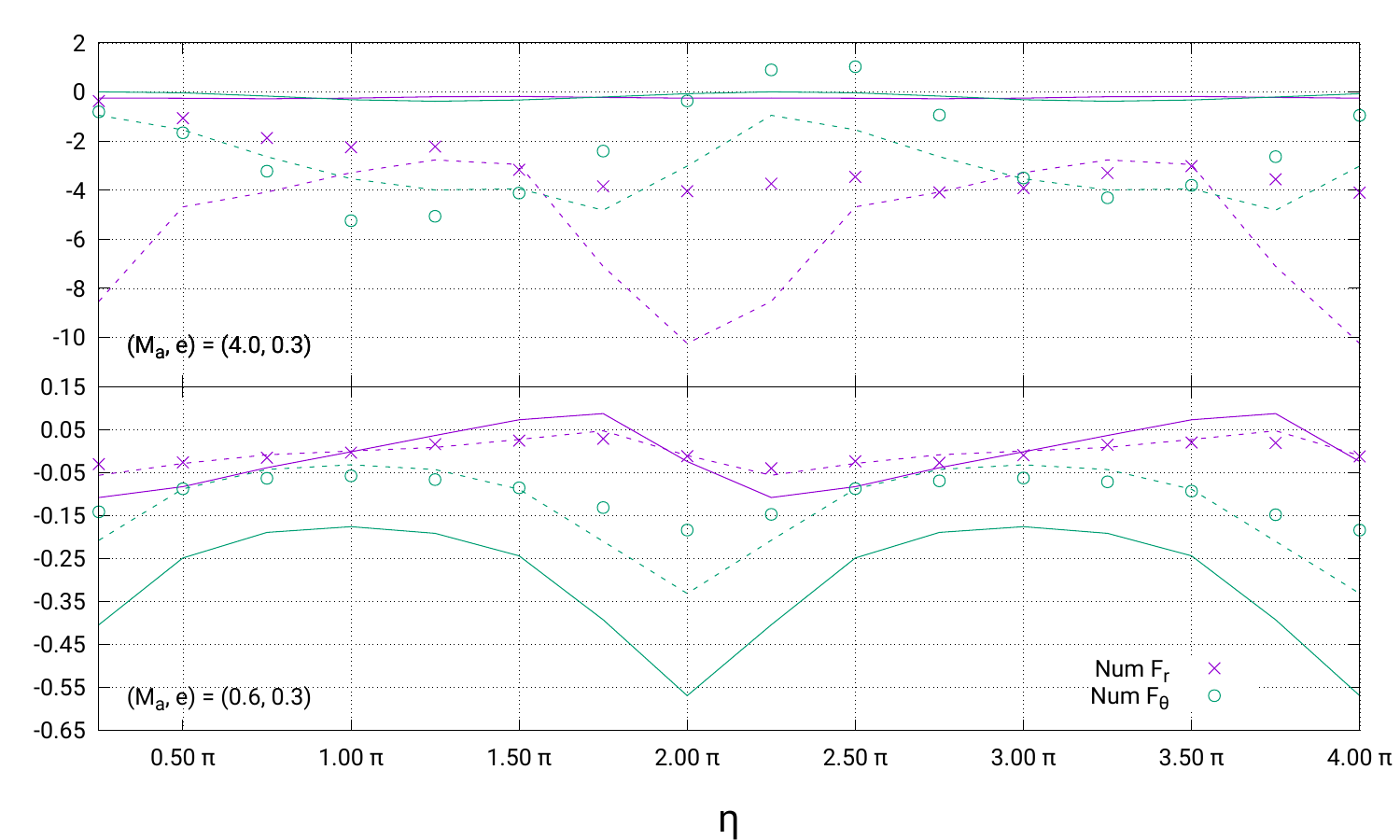}
    \caption{Radial and azimuthal components $F_r$ and $F_\vartheta$ in units of $4 \pi \bar \rho_g \mach_a^2 (G M/\Omega a)^2$ for two different parameter choices $(\mach_a, e)=(4.0,\ 0.3)$ (top panel) and $(0.6,\ 0.3)$ (bottom panel). Data points represent the DF force extracted from 3-dimensional simulations of the gas density response. The solid curve is the prediction obtained by the method outlined here, whereas the dotted curve follows from substituting the instantaneous Mach number (on the eccentric orbit) in the circular DF solution.}
    \label{fig:circ}
\end{figure}

Like the linear and circular motion case, the DF force with $e>0$ exhibits a short-distance, logarithmic divergence when the instantaneous Mach Number is supersonic, regardless of the choice of orbital parameters. This Coulomb (logarithmic) divergence is encoded in our perturbative approach and regularized with the introduction of a maximum multipole (set to match the resolution of the simulations). By contrast, the DF force always converges when the orbital velocity is locally subsonic.

We have also investigated the impact of DF on the long-time evolution of the eccentricity in the range $0.75\lesssim \mach_a\lesssim 1.05$ where our theoretical approximation is a reasonable description of the true DF force. The latter leads to orbital decay and the inspiraling of the perturber, such that the characteristic Mach number grows with time. Therefore, initial conditions are laid down at $\mach_a=0.75$ and the system is evolved until $\mach_a=1.05$.
We have checked that the time evolution of the orbit-averaged orbital parameters closely matches that obtained from a numerical integration of the instantaneous DF across 1000 orbits. The eccentric evolution depends on the initial eccentricity $e_i$ (set when $\mach_a=0.75$): for $e_i\lesssim 0.8$, the orbit tends to circularize by the time $\mach_a=1.05$ is achieved while, for $e_i\gtrsim 0.8$, it becomes more eccentric. At a qualitative level, this behaviour reflects the fact that the tangential component of the DF force can be directed along the motion near pericenter passage when the eccentricity is high. At a quantitative level, the limit between orbit circularization and eccentricity growth is reasonably predicted by comparing the relative loss of specific orbital energy and angular momentum at those orbital positions where the gain and loss of eccentricity are largest. 


Our approach, which has focused on a single perturber in an eccentric orbit, can be readily extended to a binary system along the lines of \cite{desjacques/etal:2022}. It can also include the self-gravity of the medium, be it gaseous or not.
However, extending the scope of this perturbative expansion to any (characteristic) Mach number requires that we can take into account the deformation of the orbit (from a circle to an ellipse). At a technical level, this looks challenging since this contribution implies both a time variation in the separation $r(\eta)$ between the perturber and its companion as well as a preferred direction in the orbital plane, which make the plane wave expansion (in spherical harmonics) less appealing. Alternatively, for moderate eccentricities $e\lesssim 0.5$ and outside the range $0.8\lesssim \mach_a\lesssim 1$ explored here, substituting the instantaneous Mach number of the eccentric orbit into the circular solution of \cite{desjacques/etal:2022} yields a better match to the simulation results (see Fig.~\ref{fig:circ}), but it performs worse than the perturbative approach for $\mach_a \sim 1$. 

\section*{Acknowledgements}

R.B., R.K. and V.D. acknowledge support by the Israel Science Foundation (grant no. 2562/20). 

\section*{Data Availability}

The data underlying this article will be shared on reasonable request to the corresponding author.

\bibliographystyle{mn2e}
\bibliography{references}

\onecolumn

\appendix

\section{Approximation to the Dynamical Friction for eccentric orbits}

\label{app:friction}

The complex exponential that appears in the argument of Eq.~(\ref{eq:FDF1}) can generally be expressed as
\begin{align}
\label{eq:ellipticalexp}
e^{i\vk\cdot(\vr_p(\eta)-\vr_p(\eta'))-i\omega(t(\eta)-t(\eta'))}
&= e^{-i\tilde\omega(\eta-\eta')+i\tilde\vk\cdot(\rvh_c(\eta)-\rvh_c(\eta'))}
\times e^{i\tilde\omega e(\sin\eta-\sin\eta')}
\times e^{i (1-\sqrt{1-e^2})(\sin\eta'-\sin\eta)\tilde\vk\cdot\yvh}
\end{align}
for a bound Keplerian orbit. Here, $\tilde\omega = \omega/\Omega$ and $\tilde k = a k$ are dimensionless frequency and wavenumber.

\subsection{Including timing variations}

Only the first term in the right-hand side is present when $e=0$. The second exponential factor, which differs from unity at first-order in $e$, reflects timing variations along the orbit relative to the circular case. The calculation is challenging owing to the last term, which arises at second order in eccentricity and breaks the planar symmetry. 

The DF force can be accurately predicted within linear response theory when the last factor is negligible. The comparison with numerical "simulations" shows (see Section \S\ref{sec:validation}) that this is a reasonable approximation when the characteristic Mach number $\mach_a$ is in the range $0.7\lesssim\mach_a\lesssim 1.1$. Note that $\mach_a$ is different from the orbit averaged Mach number $\overline\mach_a$, which is given by
\begin{align}
    \overline{\mach_a} &= \frac{1}{2\pi}\int_0^{2\pi}\!{\rm d}\eta\,\big(1-e\cos\eta\big)\,\mach(\eta) \\
    &= \frac{2}{\pi}\,\mach_a\,E(e^2) \nonumber \;,
\end{align}
where $\mach(\eta)$ is the instantaneous Mach number, Eq.~(\ref{eq:Machinstant}), and $E(x)$ is the complete elliptic integral.

\begin{figure}
    \begin{minipage}{.5\textwidth}
        \includegraphics[width=\textwidth]{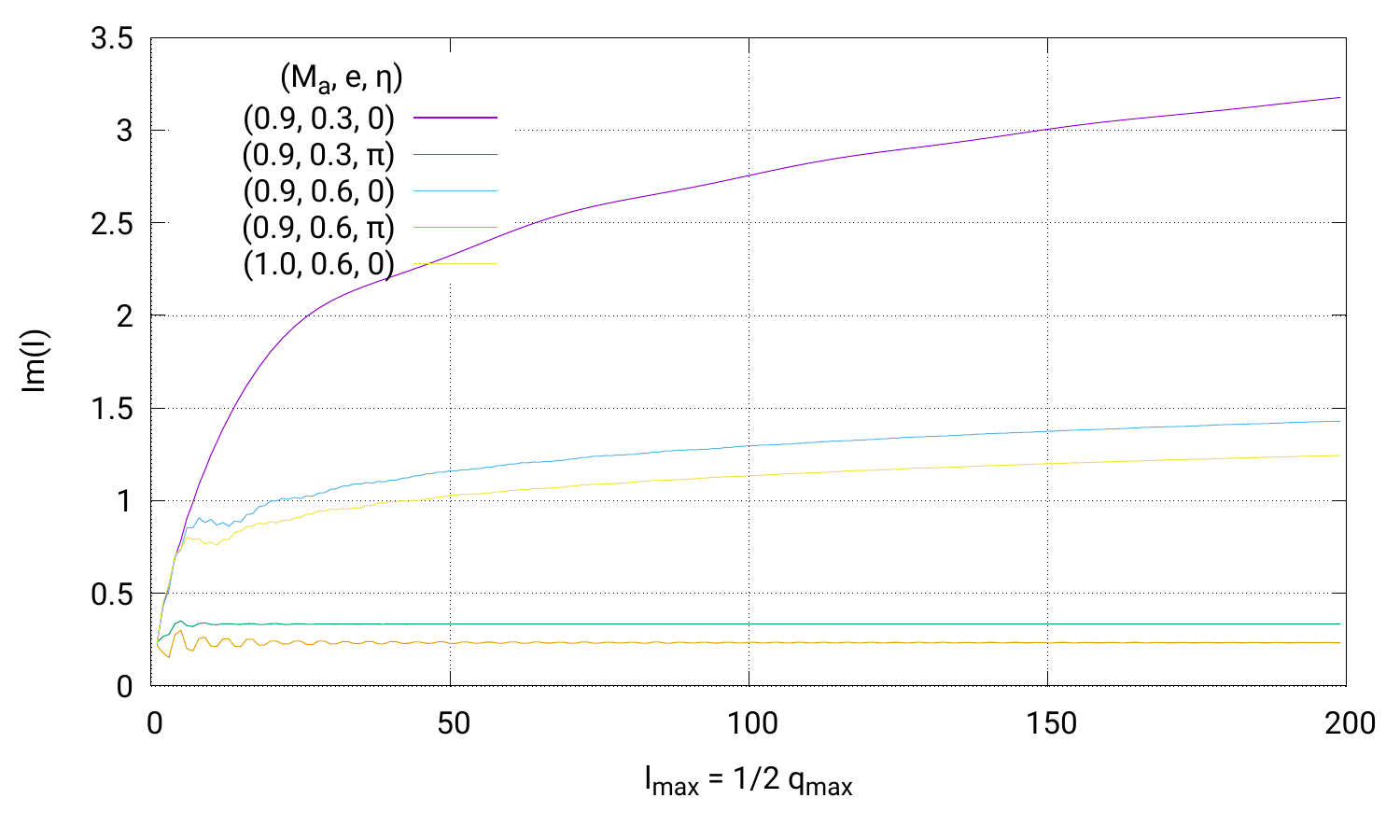}
    \end{minipage}
    \hfill
    \begin{minipage}{.5\textwidth}
        \includegraphics[width=\textwidth]{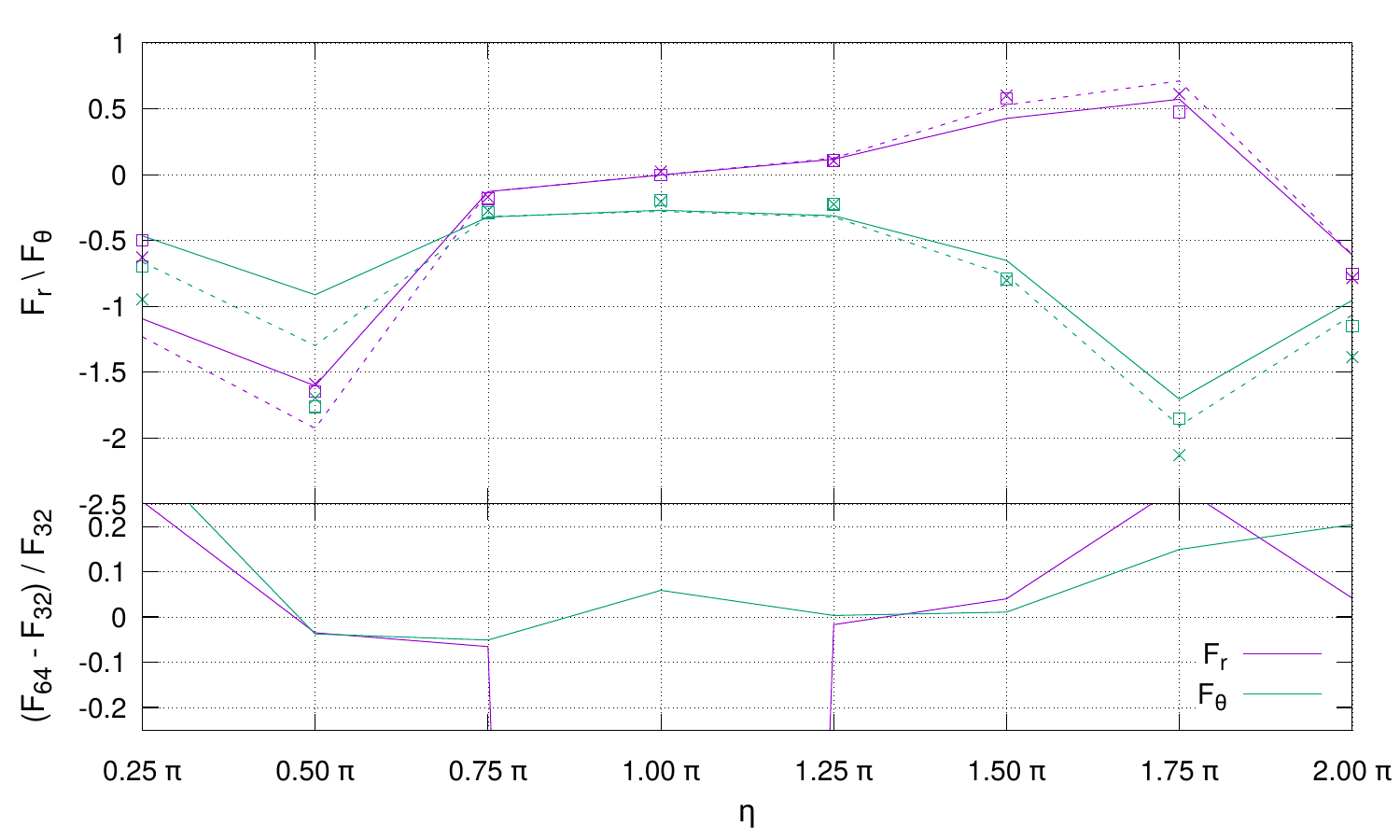}
    \end{minipage}
    \caption{{\it Left}~: Imaginary part $\Im( I(\mach_a,e,\eta))$ of the friction coefficient (Eq.~\ref{eq:Im}) for several combinations of $(\mach_a,\ e,\ \eta)$. Results are shown as a function of $l_{max}$ and $q_{max}$ such that $q_{max}=\frac{1}{2}l_{max}$ (see text for details). For all the parameter combinations corresponding to an instantaneous Mach number larger than unity, $\Im( I(\mach_a,e,\eta))$ exhibits a logarithmic divergence, whereas convergence is achieved when the instantaneous Mach number is subsonic
    {\it Right}~: The upper panel shows DF force extracted from two simulations of the linear response density $\alpha(\vr,t)$ with resolution corresponding to $\ell_\text{max}=32$ (square symbol) and 64 (cross symbol). Our theoretical predictions with $\ell_\text{max}=32$ (solid curve) and 64 (dotted curve) are overlaid for comparison. The lower panel displays the fractional difference between the two simulation results. Results are shown for $(\mach_a,\ e)=(0.9, 0.6)$.}
    \label{fig:convergence1}
\end{figure}

Ignoring the exponential involving $\vk\cdot\yvh$, we can proceed as in the circular case and use the Rayleigh expansion of  $e^{i\vk\cdot(\vr_c(\eta)-\vr_c(\eta'))}$ and the Gaunt integral to write the components $F^{(m)}(\eta)$ of the DF force in the helicity basis $(\zvh,\evh_+,\evh_-)$ as
\begin{align}
    F^{(m)}(\eta) &= 32\pi (GM)^2 \bar\rho_g \sum_{\ell_1,m_1}\sum_{\ell_2,m_2}i^{\ell_1+1}(-i)^{\ell_2}\, \sqrt{(2\ell_1+1)(2\ell_2+1)}\, Y_{\ell_1}^{m_1}\left(\frac{\pi}{2},0\right)\,Y_{\ell_2}^{m_2}\left(\frac{\pi}{2},0\right)\,\left(\begin{array}{ccc}\ell_1 & \ell_2 & 1 \\ 0 & 0 & 0 \end{array}\right) \left(\begin{array}{ccc}\ell_1 & \ell_2 & 1 \\ m_1 & m_2 & m \end{array}\right) \nonumber \\
    &\qquad \times \int_{\tilde\omega} \int_{-\infty}^{+\infty}\!\mathrm{d}\eta'\,h\big(t(\eta')\big) \big(1-e\cos\eta'\big)\,e^{-i m_1 \eta-i m_2\eta'-i\tilde\omega(\eta-\eta')}\,e^{i\tilde\omega e(\sin\eta-\sin\eta')}\, \int_0^\infty\! d\tilde k\,\frac{\tilde k j_{\ell}(\tilde k)j_{\ell-1}(\tilde k)}{\tilde k^2-(\tilde\omega\mach+i\epsilon)^2} \;,
\end{align}
with $j_l(k)$ being the spherical Bessel functions. On exploiting the symmetries of the Wigner 3j symbols, this expression can be simplified further as in \cite{desjacques/etal:2022} and we arrive at
\begin{equation}
F^{(+1)}(\eta) = 4\pi \left(\frac{GM}{\Omega a}\right)^2 \bar\rho_g \frac{e^{i\eta}}{\sqrt{2}}\,I(\mach_a,e,\eta)
\end{equation}
where the dimensionless friction coefficient $I(\mach_a,e,\eta)$ can be recast into the form
\begin{equation}
    \label{eq:Im}
    I(\mach_a,e,\eta) =\frac{\mach^2}{4}\sum_{\ell=1}^{\infty}\sum_{m=-\ell}^{\ell-2} (-1)^m\, \frac{\left(S_{\ell,\ell-1}^{(m+1)*}(\mach_a,e,\eta)-\sllm(\mach_a,e,\eta)\right)}{\Gamma\!\left(\frac{1-\ell-m}{2}\right)^2\,\Gamma\!\left(\frac{\ell-m}{2}\right)\,\Gamma\!\left(1+\frac{\ell+m}{2}\right)} \;.
\end{equation}
Here, $\Gamma(z)$ is the Gamma function while $\sllm$ is defined as
\begin{equation}
    \label{eq:scat}
     \sllm\!(\mach_a,e,\eta)=\int_{\tilde\omega} \int_{-\infty}^{+\infty}\!\mathrm{d}\eta'\,h\big(t(\eta')\big)\,\big(1-e\cos\eta'\big)\,e^{i(\tilde\omega-m)(\eta'-\eta)+i \tilde \omega e(\sin(\eta)-\sin(\eta'))}\int_0^\infty\! d\tilde k\,\frac{\tilde k j_{\ell}(\tilde k)j_{\ell-1}(\tilde k)}{\tilde k^2-(\tilde\omega\mach_a+i\epsilon)^2}\;.
\end{equation}
In steady state, $h(t)=1$ and the integral over $\eta'$ can be carried out using the Jacobi-Anger relation
\begin{align}
    e^{-i \tilde \omega e \sin (\eta')}&=\sum_{q=-\infty}^\infty J_{q}(\tilde \omega e)e^{-i q\eta'}\\
        &=J_0(\tilde\omega e)+J_1(\tilde\omega e)(e^{-i\eta'}-e^{i\eta'}) + J_2(\tilde\omega e)(e^{-i2\eta'}+e^{i2\eta'})+\dots
\end{align} 
with $J_q(k)$ being the cylindrical Bessel functions. This leads to expressions of the form
\begin{align}
    \int_{-\infty}^{\infty}\mathrm d\eta' e^{i \alpha (\eta'-\eta)}e^{\pm i n\eta'}(1-e \cos\eta')&=2\pi \left[e^{\pm i n\eta}\del{\alpha\pm n}-\frac{e}{2}\left( e^{\pm i(n\pm 1)\eta}\del{\alpha\pm n+1}+e^{-i(n\mp 1)\eta}\del{\alpha\pm n-1}\right) \right]
\end{align}
after taking advantage of $\int_{-\infty}^{\infty}\mathrm d x\ e^{ik x}=2\pi\, \del{k}$. Substituting these relations into Eq.~(\ref{eq:scat}) and rearranging the terms, we arrive at
\begin{align}
     S_{\ell,\ell-1}^m\!(\mach_a,e,\eta)&=2\pi\int_{\tilde\omega} e^{i\tilde \omega e \sin(\eta)}\,\bigg\{\del{\alpha}\bessJb{0}{\tw e}\\
     &+\del{\alpha-1}e^{-i\eta} \bigg[\bessJb{1}{\tw e}-\frac{e}{2}\Big(\bessJb{0}{\tw e}+\bessJb{2}{\tw e}\Big)\bigg]+\del{\alpha+1}e^{i\eta} \bigg[-\bessJb{1}{\tw e}-\frac{e}{2}\Big(\bessJb{0}{\tw e}+\bessJb{2}{\tw e}\Big)\bigg]
     \nonumber \\
     &+\del{\alpha-2}e^{-i2\eta} \bigg[\bessJb{2}{\tw e}-\frac{e}{2}\Big(\bessJb{1}{\tw e}+\bessJb{3}{\tw e}\Big)\bigg]+\del{\alpha+2}e^{i2\eta} \bigg[\bessJb{2}{\tw e}+\frac{e}{2}(\bessJb{1}{\tw e}+\bessJb{3}{\tw e})\bigg] \nonumber \\
     &+\dots\bigg\}
     \cdot \int_0^\infty\! d\tilde k\,\frac{\tilde k j_{\ell}(\tilde k)j_{\ell-1}(\tilde k)}{\tilde k^2-(\tilde\omega\mach_a+i\epsilon)^2} \nonumber\;,
\end{align}
where $\alpha=\tw-m$ and we have included contributions proportional to $\bessJb{3}{\tw e}$ for completeness. The $\tw$-integral can now be easily carried out. Furthermore, the coefficients of the terms proportional to $\del{\alpha-q} e^{-iq\eta}$ exhibit a similar structure which we can easily work out with aid of the Jacobi-Anger relation. We find the general formula
\begin{align}
    A_q(e,m,\eta)=e^{i(e(m+q)\sin\eta-q\eta)}\cdot \left \lbrace\begin{array}{ll}
        J_0(me) \Big. &q=0  \\
        -J_{|q|}(e(m+q))-\frac{e}{2}\Big[J_{|q|-1}\big(e(m+q)\big) +J_{|q|+1}\big(e(m+q)\big) \Big] & q\textit{ odd}\ \&\ q<0\\
        J_{|q|}\big(e(m+q)\big)+\frac{e}{2}\Big[J_{|q|-1}\big(e(m+q)\big) +J_{|q|+1}(e(m+q)) \Big] & q\textit{ even}\ \&\ q<0\\
         J_{|q|}\big(e(m+q)\big)-\frac{e}{2}\Big[J_{|q|-1}\big(e(m+q)\big) +J_{|q|+1}\big(e(m+q)\big) \Big] & \textit{otherwise}\\
    \end{array} \right.
\end{align}
and write
\begin{equation*}
      S_{\ell,\ell-1}^m\!(\mach_a,e,\eta)=\sum_{q=-\infty}^{\infty} A_q(e,m,\eta) \int_0^\infty\! d\tilde k\,\frac{\tilde k j_{\ell}(\tilde k)j_{\ell-1}(\tilde k)}{\tilde k^2-((m+q)\mach_a+i\epsilon)^2}\;.
\end{equation*}
The remaining $\tilde k$-integral is identical to its circular counterpart (with $m$ replaced by $m+q$), which we solved in \cite{desjacques/etal:2022} (see their equations (17) and (18)): 
\begin{align}
    S_{\ell,\ell-1}^{m}(\mach_a)=\left\lbrace \begin{array}{ll}\frac{i \pi}{2} j_\ell\big(m\mach_a\big)\, h_{\ell-1}^{(1)}\!\big(m\mach_a\big)&m\neq 0\\
        \frac{\pi}{2(4\ell^2-1)}& m=0\end{array} \right. \;.
\end{align}
The Hankel function $h^{(1)}_\ell$ represents an outgoing wave, which follows from the choice of the retarded Green's function.

The final result thus takes the compact form
\begin{equation}
    \label{eq:SMetae}
    S_{\ell,\ell-1}^m(\mach_a,e,\eta)=\sum_{q=-\infty}^{\infty} A_q(e,m,\eta)\,  S_{\ell,\ell-1}^{m+q}(\mach_a)\;.
\end{equation}
Note that, at small eccentricity $e\ll 1$, the coefficients $A_q(e,m,\eta)$ scale like $e^q$.

\begin{figure}
    \begin{minipage}{.5\textwidth}
        \includegraphics[width=\textwidth]{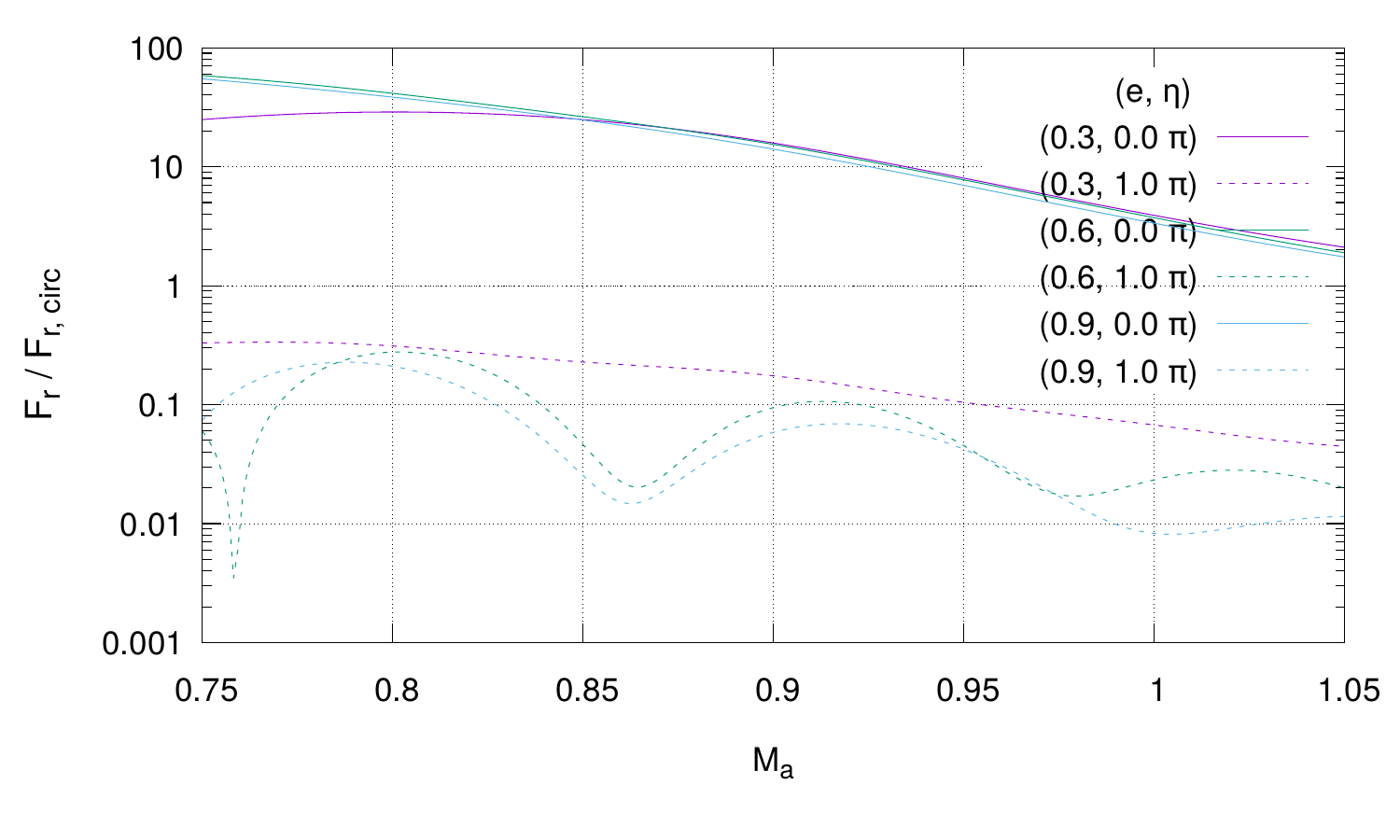}
    \end{minipage}
    \hfill
    \begin{minipage}{.5\textwidth}
        \includegraphics[width=\textwidth]{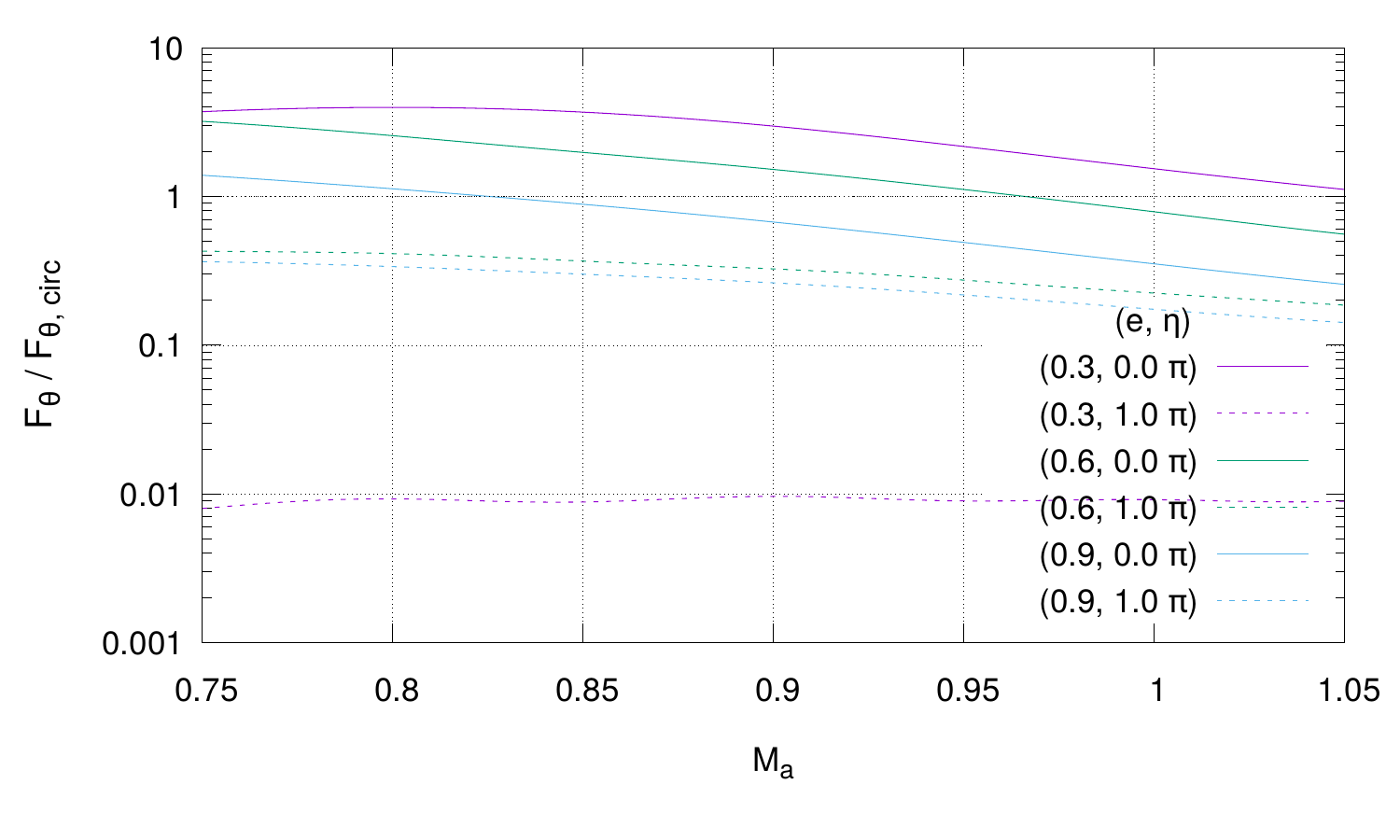}
    \end{minipage}
    \caption{{\it Left panel}: Ratio $F_r/F_{r, \mathrm{circ}}$ of the eccentric and circular radial DF component at pericenter and apocenter passage for different eccentricities. Results are shown in the range of characteristic Mach number $\mach_a$ where the agreement between our approximation and the simulation is best. The horizontal dotted black lines indicates $F_r/F_{r, \mathrm{circ}}=1$. {\it Right panel}: Same as the left panel but for the tangential component of the DF force.}
    \label{fig:df_elip_circ}
\end{figure}

\subsection{Coulomb divergence}

In order to evaluate numerically the friction coefficient $I(\mach_a,e,\eta)$ given by Eq.~(\ref{eq:Im}), it is necessary to truncate the multipole expansion at some finite $\ell_\mathrm{max}$ and introduce a upper (resp. lower) cut-off $q_\mathrm{max}$ (resp. $-q_\mathrm{max}$) in the summation in Eq.~(\ref{eq:SMetae}). 

In the left panel of Fig.~\ref{fig:convergence1}, we investigate the convergence of the friction coefficient at periapsis and apoapsis for various parameter combinations $(\mach_a, e, \eta)$.
We vary $l_{max}$ and $q_{max}$ simultaneously according to the empirical relation $q_{max}=\frac{1}{2}l_{max}$ which, for a given $l_{max}$, determines the range of $q$ beyond which the sum converges. When the instantaneous Mach number (Eq.~\ref{eq:Machinstant}) is subsonic, the multipole expansion quickly converges. By contrast, the friction coefficient exhibits a logarithmic divergence similar to the circular case $e=0$ \citep[see Fig. 3 in][]{desjacques/etal:2022} when the instantaneous motion is supersonic. We also checked that the radial component converges for all the choices of $(\mach_a, e, \eta)$ considered here.

In the right panel of Fig.~\ref{fig:convergence1}, we compare our theoretical prediction with the DF force extracted from two 3-dimensional simulations of the linear response density $\alpha(\vr,t)$ (see \S\ref{sec:validation}), the first with a mesh resolution $\Delta=a/32$ and the second with $\Delta=a/64$. Although both simulations agree on the DF force around apocenter passage, the higher resolution simulation yields a larger DF force around pericenter passage, where the instantaneous Mach Number is largest. Our theoretical predictions, which assume $(l_\mathrm{max},q_\mathrm{max})=(30,15)$ and $(60,30)$ for the low and high simulations respectively, reproduce the amplitude of this effect. All this suggests that the short distance Coulomb divergence is present as soon as the instantaneous Mach number on the eccentric orbit is supersonic. 

\subsection{Comparison with the circular result}

In Fig.~\ref{fig:df_elip_circ}, we compare our approximation to the eccentric DF force to the circular solution of \cite{desjacques/etal:2022}. Results are shown for the radial (left panel) and tangential (right panel) component $F_r$ and $F_\vartheta$ across the range of characteristic Mach number $0.75\lesssim\mach_a\lesssim 1.05$ where our approximation matches best the simulation results. 

At pericenter, $F_r$ exceeds the circular expectation by a factor as large as $\approx 60$ at $\mach_a\sim 0.8$ before it drops off towards larger Mach-numbers. This behaviour stems from the large increase in the radial component as the instantaneous Mach number approaches the transition to supersonic motion. Such an enhancement can occur at relatively low values of $\mach_a$ when the orbit is highly eccentric.
Conversely, the eccentric DF force is always smaller than its circular counterpart at apocenter, where the instantaneous Mach number can be much smaller than $\mach_a$ when $e\sim 1$. The origin of the oscillatory pattern which emerges at high eccentricity is unknown, although we suspect that it is a numerical artifact. 
At pericenter, the tangential component shows a behaviour similar to the radial part, although the enhancement relative to the circular case does not exceed $\approx 4$ for $\mach_a\sim 0.8$. Furthermore, $F_\vartheta$ can be smaller than its circular counterpart at higher $\mach_a$ depending on the eccentricity. At apocenter however, the eccentric $F_\vartheta$ always falls below the circular expectation without a clear trend with eccentricity. The relative suppression is strongest for $e=0.3$ and smallest for $e=0.6$, before it increases again towards $e=0.9$. 



Even if variations in the instantaneous Mach number can explain part of the observed trend, it is not sufficient to explain the detailed behaviour of the eccentric DF force, especially the structure of maxima and minima seen in $F_r$ at high eccentricities.

\label{lastpage}

\end{document}